FERMILAB-FN-1128-TD

# PIC Simulations of One-side Multipactor on Dielectric

The study submitted to 2021 International Conference on RF Superconductivity (SRF'21)

Gennady Romanov

## Abstract

Breakdown of dielectric RF windows is an important issue for particle accelerators and high-power RF sources. One of the generally considered reasons for the RF windows failure is the multipactor on dielectric surface. The multipactor may be responsible for excessive heating of dielectric and discharge of charges that accumulated in ceramic due to secondary emission. In this study the comprehensive self-consistent PIC simulations with space charge effect were performed in order to better understand the dynamic of one-side multipactor development and floating potential on dielectric induced by the emission. The important correlations between the multipactor parameters at saturation and the secondary emission properties of dielectric and the applied RF field parameters were found and are reported in the paper.

## Introduction

One side multipactor, which is typical for RF windows, requires a returning force to develop. In case of isolated metal or dielectric body the returning force can be a result of floating potential which is due to charging of the isolated body by emission current. Also, an inhomogeneous RF field can by itself ensure the return of the emitted electrons to the body surface, but this case is not considered here. Buildup in the surface charge starts with random colliding electrons that come from other processes and sources with energy enough to generate larger number of secondary electrons. If the certain conditions are met, then, at early stage of multipactor development, the emission current (the secondary electrons that leave the body) is larger than the collision current (the electrons that return to and hit the body), so the surface charge buildup continues, and positive electric charge is accumulated on the body. With increasing of the returning force more and more of the secondary electrons start to return to the emitting surface and contribute to the floating potential. This stochastic process requires sufficiently high secondary emission yield of material (SEY) to be realized, and, unfortunately, the dielectric materials of RF windows typically have very high secondary emission yield (SEY=8-10 for alumina). Obviously that this charging cannot continue indefinitely and eventually the process comes to saturation at some equilibrium floating potential on dielectric.

The time-dependent physics of the one-side multipactor was studied in detail with self-consistent particle in cell (PIC) numerical simulations using CST Particle Studio. The main advantages of this PIC solver are true multiparticle dynamic, 3D space charge distribution, RF and static fields distortion due to the space charge impact and the surrounding, advanced secondary emission models. It turned out that the realistic energy spread of the secondary electrons to a large extent defines the dynamic of this type multipactor.

## Particle-in-Cell model

The principal PIC model is simple: it is a dielectric plate placed in the static and radiofrequency (RF) electric fields. Uniform electrostatic electric field is perpendicular to the plate surface and acts as a returning force in the simulations without space charge effect, and it is disabled in simulations with space charge effects. Uniform RF electric field is parallel to the dielectric surface (Fig.1) and provides the electrons with energy for generation the secondary electrons. The equations of the electron motion in this case are as follows:

$$m\ddot{y} = -eE_{DC}; \quad m\ddot{x} = -eE_{rf0}\sin(2\pi ft + \theta), \quad (1)$$

where $x$ and $y$ are respectively horizontal and vertical coordinate of the electron; $m$ – electron mass; $e$ – electron charge; $E_{DC}$ – static electric field; $E_{rf0}$ – amplitude of RF electric field; $f$ – frequency of the RF field; $\theta$ – phase of the RF field at the moment of electron emission (initial phase of the emitted particle).

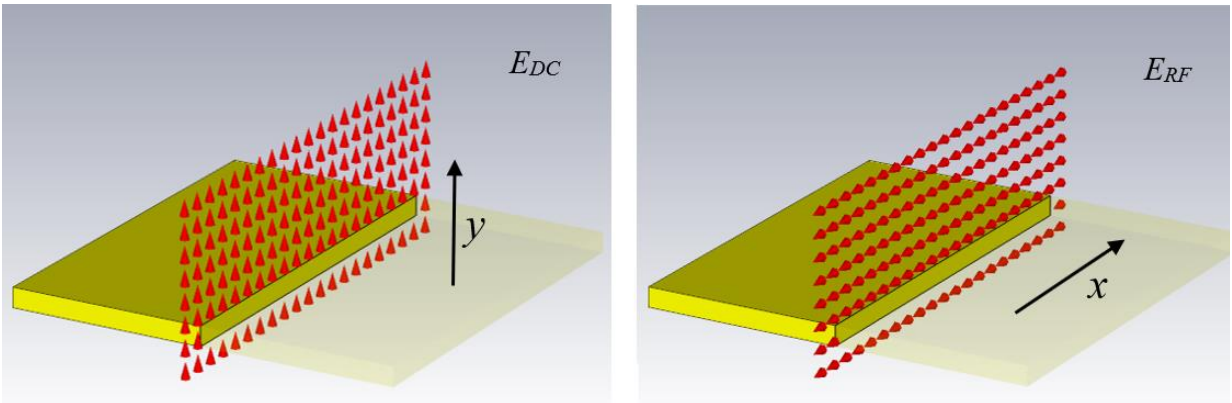

Figure1: Cross-sections of electrostatic and RF filed distributions.

The emission property of plate's material is provided by assigned secondary emission model. The advanced probabilistic Furman emission model from CST library, which includes elastic and diffusion emissions, was not used in this PIC model, since the simulations were performed mostly with GPU acceleration, which works only with imported true emissions. Besides there are no reliable data on elastic and diffusion emissions for RF ceramics. Because of these two reasons it was decided not to use these kinds of emission at all, and the dielectric plate was provided with the imported Vaughan emission model, general SEY function of which is shown in Fig.2. The maximums of SEY functions varied from 1.5 to 3, which is much lower than a real emission of dielectrics can be. The SEY was lowered in the simulations to avoid excessive number of particles being tracked and reduce time of simulation.

Random gamma distributed initial energy of secondary electrons $W_0$ with the most probable energy of 7.5 eV was used in all simulations. The probability density function of the distribution is also shown in Fig.2.

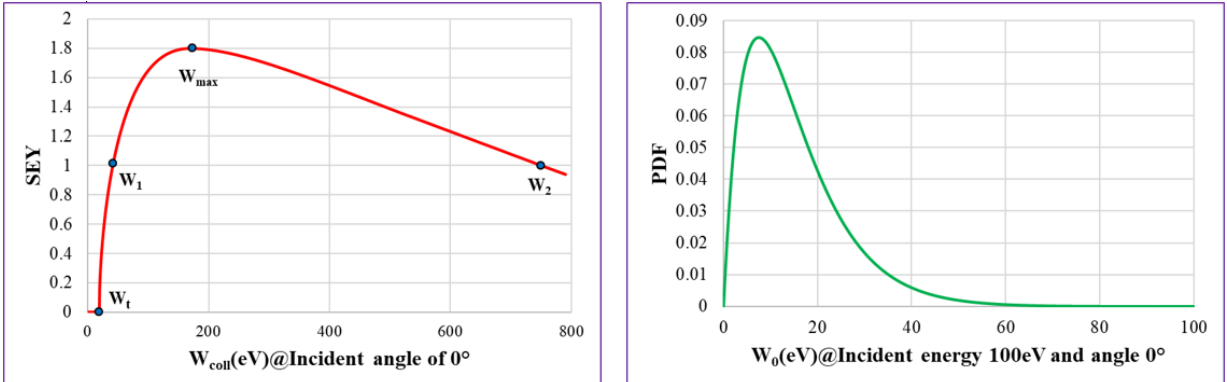

Figure 2: The general Vaughan SEY function and the probability density function of initial energy of secondary electrons. In the SEY plot the important incident energies are marked: threshold energy $W_t$, first crossover $W_1$, SEY maximal $W_{max}$ and second crossover $W_2$.

The source of initial particles was placed in the center of the plate. It emitted particles at start of the simulations during one RF period $T=1/f$ to cover all possible initial phases of particles. For easier interpretation of starting stage of simulation, the initial electrons were monoenergetic with fixed energy of 7.5 eV and did not have angular spread – they all were emitted perpendicularly to the surface. Note, that this setting worked for initial particles only – during further simulations the parameters of secondary electrons were governed by chosen emission model.

## Simulations without space charge effect

These simulations were performed to evaluate the field levels that are favorable for multipactor development. The analytically estimated threshold fields were $E_{RF}$ = 21.13 kV/m for 325 MHz RF field (amplitude, which provides maximal parallel to the plate acceleration up to 45 eV to a secondary electron during half RF period and launched at initial phase $\theta=0°$), and $E_{DC}$ = 12 kV/m for the electrostatic field (provides time of flight equal to half of RF period $T/2$ for secondary electron with most probable emission energy of 7.5 eV). In the simulations the RF field amplitude was swept from 18 kV/m to 110 kV/m, and the electrostatic field was changed from 7 kV/m to 26 kV/m.

The parameters of the used emission model were maximal $SEY_{max}=1.8$ at $W_{max}=150$ eV, $W_t=0$, $W_1=22$ eV and $W_2=1147$ eV.

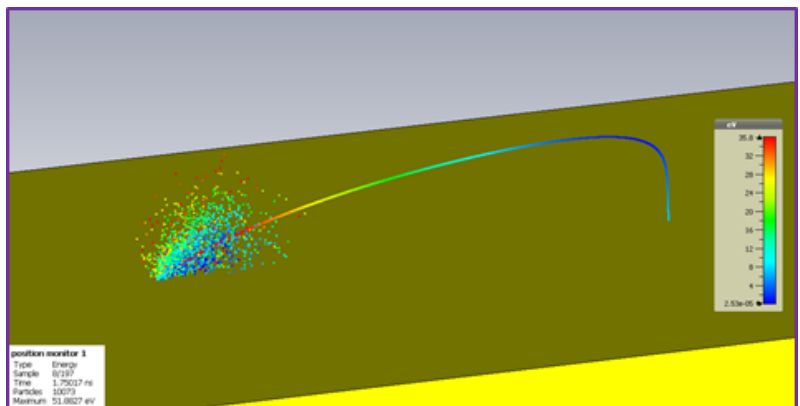

Figure 3: Particle distribution at 1.75 ns after start of emission. The curve before collision is not a particle trajectory, but a continuous chain of particles. After the collision there is a cloud of the secondary electrons with random initial energies and directions.

In these simulations without space charge effects the single point particle source was used for clearer picture of emission process. In Fig.3 the evolution of the emitted beam in the crossed analytically estimated fields during half of RF period is shown. The head of the electron train is emitted at $t = 0$ and initial phase $\theta = 0$ and hits the plate at t = T/2. The collision energy of the electrons is appropriate to generate a bunch of the secondary electrons, which have now different velocities by values and directions accordingly to the emission model.

Further developments of MP simulated at different $E_{DC}$ are shown in Fig.4. The simulations show that MP starts slightly earlier than analytical estimation of 12 kV/m, but on the other hand at slightly higher RF field amplitude of 24 kV/m. At $E_{RF}$ < 24 kV/m MP didn't start at any level of the electrostatic field. Noticeable, that number of particles vs time demonstrates exponential growth and resonance character.

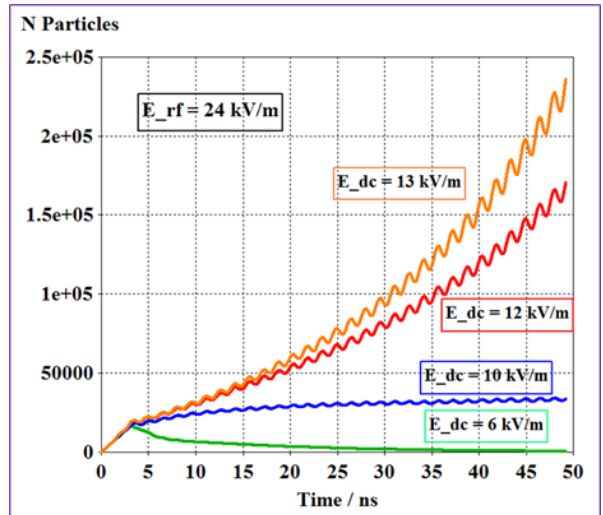

Figure 4: Number of particles vs time at different returning electrostatic field strength and RF field amplitude of 24 kV/m.

More statistical data were collected with the emission parameters $SEY_{max}$=1.5 at $W_{max}$=150 eV, $W_t$=0, $W_1$=33.5 eV and $W_2$=667 eV. Also, the model itself got some minor improvements like enhanced mesh and reduced "leak" current among them.

The data are presented in Fig.5-7. Simulation time was 15 RF periods, this relatively short time was chosen to reduce overall time of simulations. It created not a correct situation when a collision current I_coll ≠0 even at <SEY> < 1 (the macroparticles have electric charge and induced current can be calculated, though the space charge effect is not considered). The particles from initial bunch do not have enough time to disappear and continue colliding with the plate. So, it should be kept in mind that below <SEY>=1 the collision current I_coll and collision energy W_coll go to zero after enough time. The averaging of the parameters was performed over last 5 RF periods.

The results of the simulations without space charge effect are not realistic but help to define the range of parameters and to understand the correlations between them. In particular let mark that MP process starts (<SEY> exceeds unit at first time) at $E_{rf}$ = 29.05 kV/m and $E_{dc}$ = 13.5 kV/m. $E_{rf}$ =29 kV/m is a lowest level of RF field at which multiplication starts and it again exceeds the analytical estimation of 17 kV/m. Apparently, there are not enough particles accelerated by the field below 29 kV/m up to energies above $W_1$ that provide SEY>1 to support multiplication. The average energy of collision at this point is 37.5 eV which is slightly greater $W_1$ = 33.5 eV of the material. Electrostatic field of 13.5 kV/m makes resonant the particles with initial energy

$$W_0 = \frac{m}{2e}\left(\frac{E_{dc}e}{4fm}\right)^2 = 9.48 \ eV, \qquad (2)$$

which means that the secondary electrons with initial energy 9.5 eV (close to the most probable value of 7.5 eV and therefore the most numerous initial electrons) are resonant, having time of flight equal to the half of RF period. In general (2) indicates, that there are always the resonance secondary particles emitted with proper initial velocity at any level of electrostatic field, though the number of them after emission is different according to the PDF of initial energies as shown in Fig 2. It should be also noted that the resonant particles with initial energy that exceeds 7.5 eV significantly can exist only with an external fixed electrostatic field. The number of emitted resonant particles drops above 7.5 eV according to the given PDF. The electrostatic field due to the charge induced by emission decreases with drop of emission current and the resonance returns to the particles with initial energy 7.5 eV. An autoregulation of the positive charge on the ceramic in the similar practical case is also described in [1].

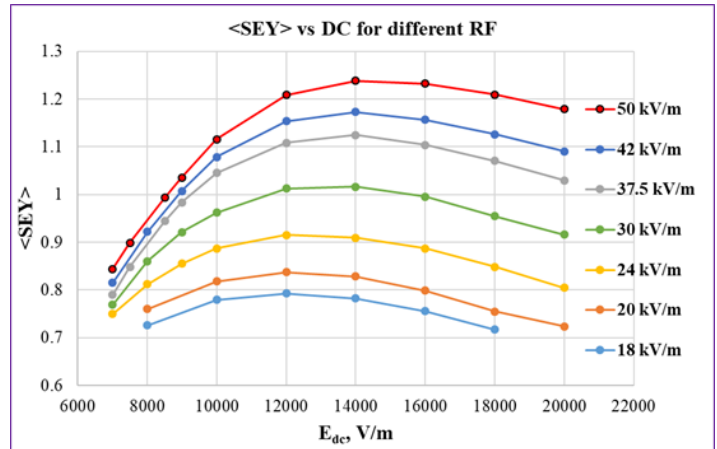

Figure 5: Average secondary emission yield <SEY> vs electrostatic field strength at different amplitudes of RF field.

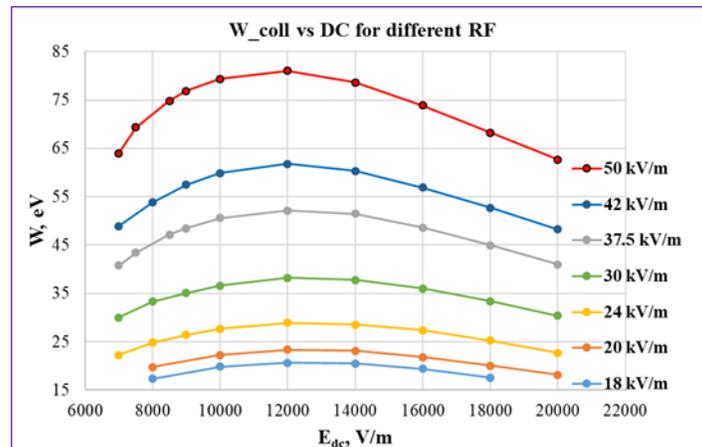

Figure 6: Average energy of the particles at the moment of collision with the dielectric plate vs electrostatic field strength at different amplitudes of RF field.

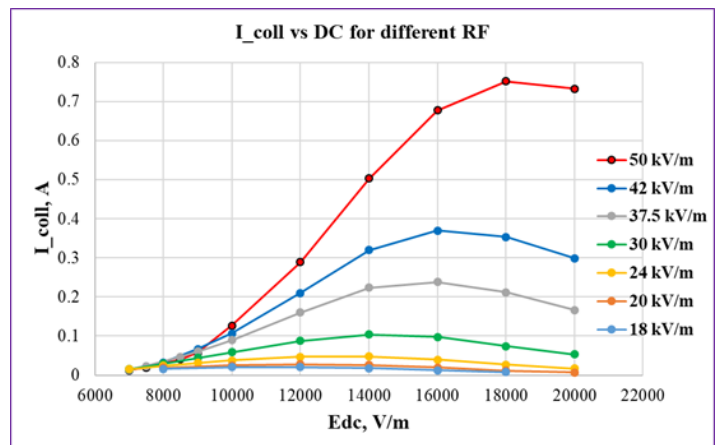

Figure 7: Average current of the particles colliding with the dielectric plate vs electrostatic field strength at different amplitudes of RF field.

Evaluation of the breakdown level of RF field for very low electrostatic field made in [2] assumes only non-resonant motion of the electrons (so called polyphase regime: time of flight τ >> T for all electrons, collision phases are uniformly distributed over RF period). This approach is correct for low electrostatic fields since the number of resonant electrons with very low initial energy is negligibly small. For the considered emission parameters $W_1$ = 33.5 eV and f=325 MHz the breakdown level of $E_{rf}$ following [2] is

$$E_{rf\_breakdown} = 0.94 \cdot 2\pi f \sqrt{\frac{2U_1 m}{e}} = 37.5\,\frac{kV}{m}, \qquad (3)$$

where $U_1 = W_1/e$ = 33.5 V is the first crossover potential. The breakdown level of $E_{rf}$ obtained in the simulations is much lower (29 kV/m), which suggests a contribution from more effective and fast resonant multiplication.

## MP saturation with space charge effect

The main impact of the space charge on virtually all kinds of multipactor is a saturation of the multipactor. In case of RF electric field parallel to dielectric surface the saturation of the discharge is combined with the saturation of the charge accumulated in dielectric. To study this time dependent process the PIC simulations were performed with active space charge effect.

For simulations with space charge effect the model was modified. The external electrostatic field was removed, a voltage monitor was added as shown in Fig.8. The single point particle source was replaced with a circular one to make an initial charging of the ceramic more uniform. Total charge emitted during one RF period was chosen equal to 1e-9 C. Vaughan emission model parameters were Wt=0, $W_1$=11 eV, maximal SEY of 3.0 at Wmax=200 eV and $W_2$=6470 eV.

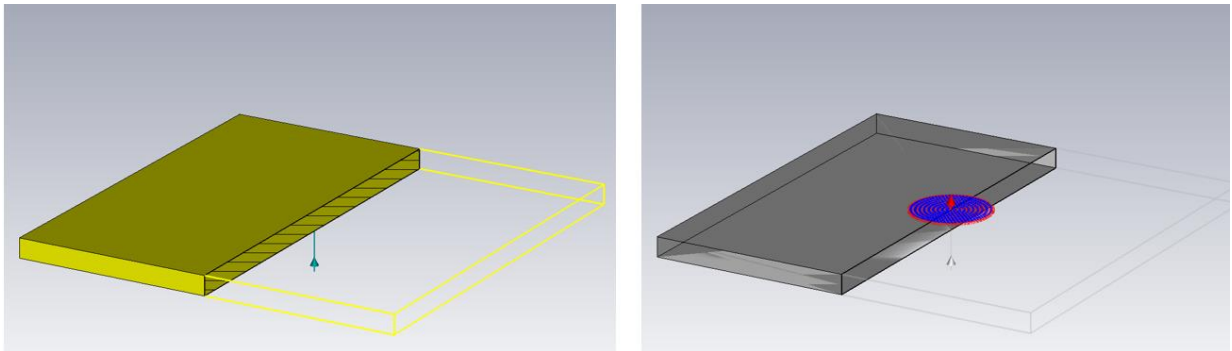

Figure 8: Left – location of the voltage monitor indicated by arrow. Right –the circular source of initial particles.

Initial emission from particle source instantly generates a potential on the dielectric surface, so there was no need to use any ancillary electrostatic field, which was used in some models to initiate multipactor process [3]. Particle distribution in space after 2 ns of emission is shown in Fig.9. Some particles leave the dielectric along straight trajectories. Apparently, they are the very first particles emitted when the electrostatic field is not enough yet to return them to the surface. Gradually the emitting electrons build up a positive charge on the dielectric and the particles start to return to the surface.

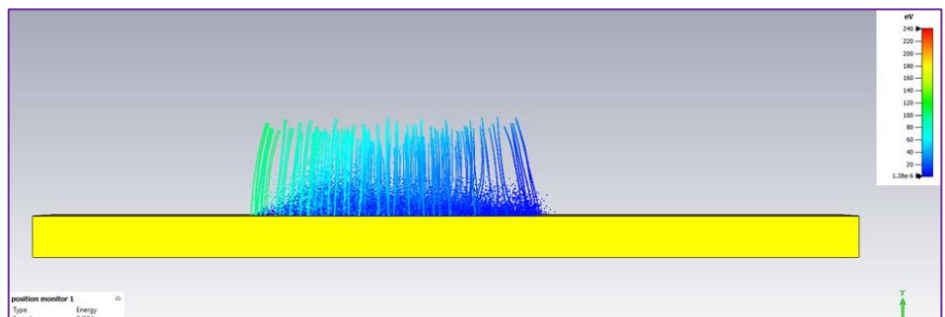

Figure 9: Particle distribution at 2 ns after start of emission.

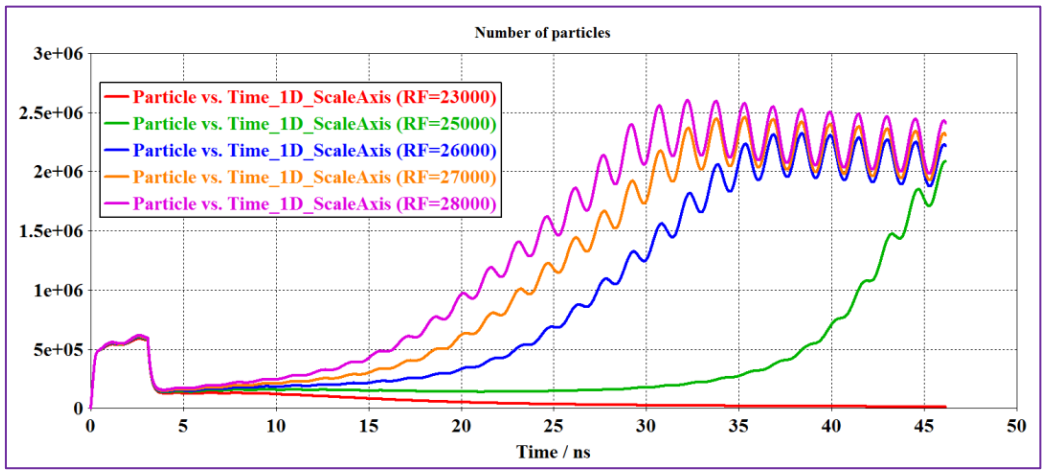

Figure 10: Number of particles vs time at different levels of RF field.

Development of MP at different levels of RF field during simulation time of 15 RF periods is shown in Fig.10. Breakdown level of $E_{RF}$ is about 25 kV/m, which is slightly lower that was found in the simulations without space charge effect with external electrostatic field. Further increasing of RF field above the breakdown level changes the speed of MP development, but the saturation level of number of particles remains almost the same at each RF field value, it just slightly increases in average.

The electrostatic field induced by MP at the end of simulation is shown in Fig. 11. The field is not uniform, and its distribution depends also on the surrounding. The voltage monitor of the electrostatic field is located under the plate and integrates electrostatic field along 2 mm line perpendicular to the plate. The location has been chosen to avoid interference of the monitor with the space charge of the particle cloud. The field strength is obviously different above and below the plate, so the monitor readings are relative. The voltage monitor in Fig.12 also shows some dependence of the saturated electrostatic field level as well as growth rate on the applied RF field. When the RF field is below the breakdown level and the plate is not charging/discharging, the constant electrostatic field shown in Fig.12 is a remnant of the charge left by the emission of the initial particles.

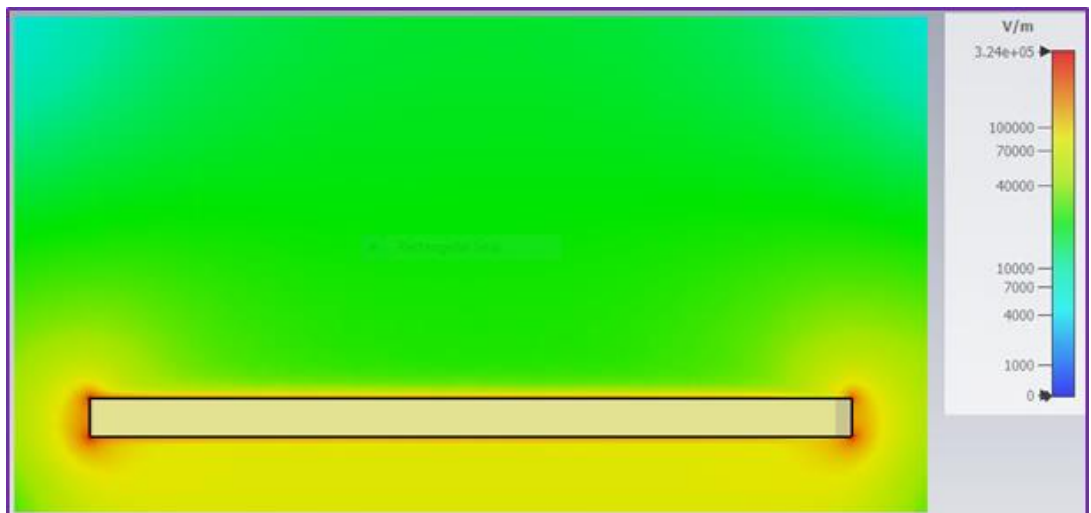

Figure 11: Electrostatic field induced by saturated multipactor process.

The collision energy vs time also saturates in the similar fashion as other MP parameters (see Fig.13). But there is one more important feature in addition to the dependence of the collision energy on the applied RF field. Namely, the phase of collision also depends on the applied RF field strength, which is shown clearly in the insert of Fig 13. This dependence will be discussed later.

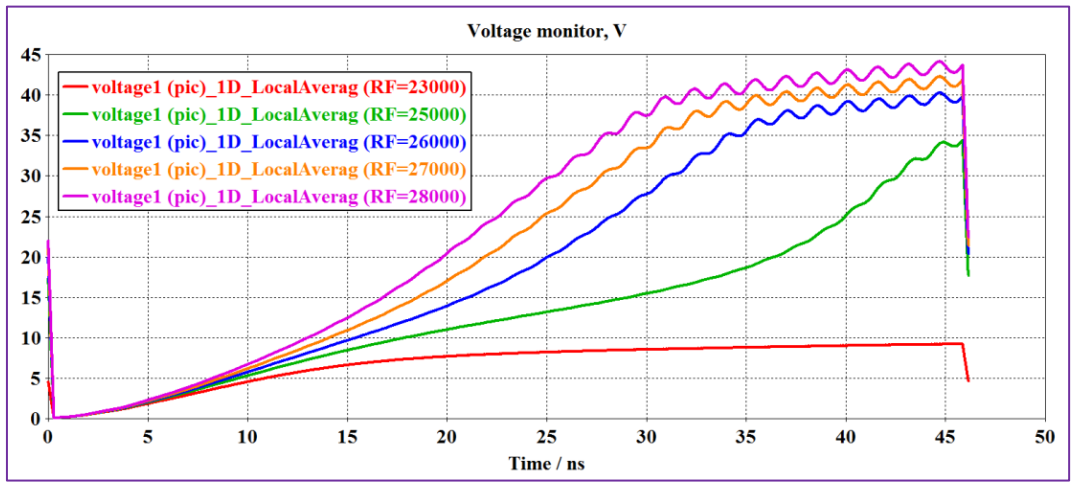

Figure 12: Data from voltage monitor for different RF field levels.

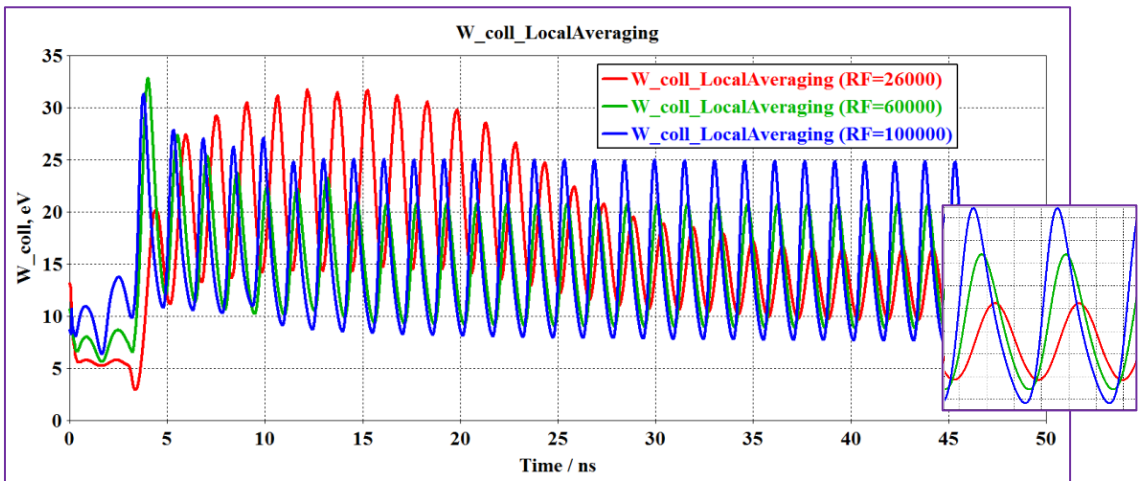

Figure 13: Collision energy vs time for different RF field levels. The insert shows the shift of collision phase with increasing of RF field level.

## Impact of the emission model parameters on MP process.

Four different SEY functions were used to investigate impact of SEY on multipactor dynamic. Their parameters are shown in the Tab.1. The initial energy distribution of the secondary particles was the same in all simulations (PDF is shown in Fig.2).

Among the emission model parameters, the first crossover $W_1$ of the SEY function plays especially important role in the MP process. It defines RF field level at which multipactor begins (threshold) and influences the saturation levels of multipactor parameters.

Table 1

| $SEY_{max}$ | $W_{max}$, eV | $W_1$, eV | $W_2$, keV |
|---|---|---|---|
| 3 | 200 | 11 | 6.6 |
| 3 | 200 | 16 | 6.5 |
| 1.8 | 150 | 22 | 1.1 |
| 3 | 220 | 31 | 7.0 |

Figure 14 shows a typical change of collision current and collision energy for different first crossovers.

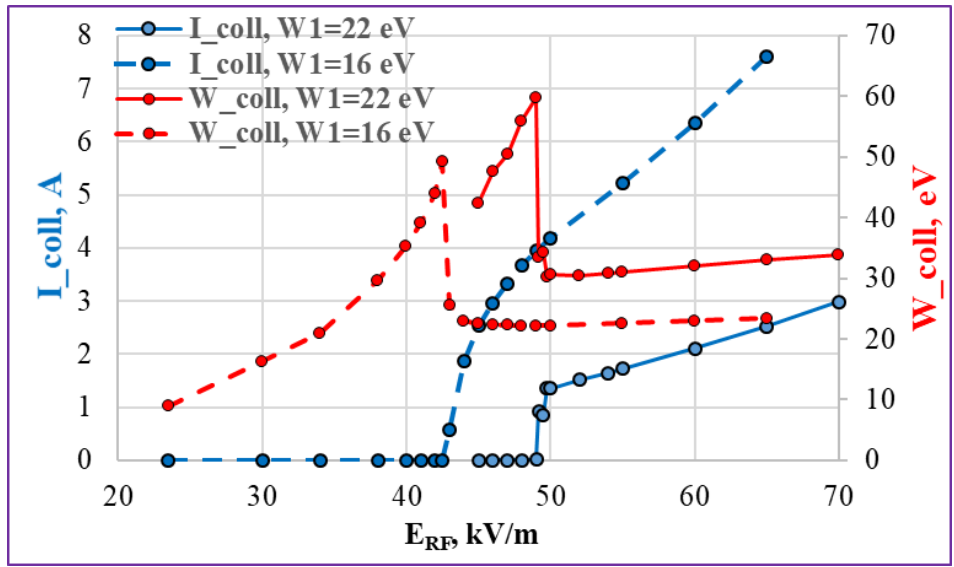

Figure 14: Average collision current I_coll and collision energy W_coll of multipactor as functions of the RF field amplitude.

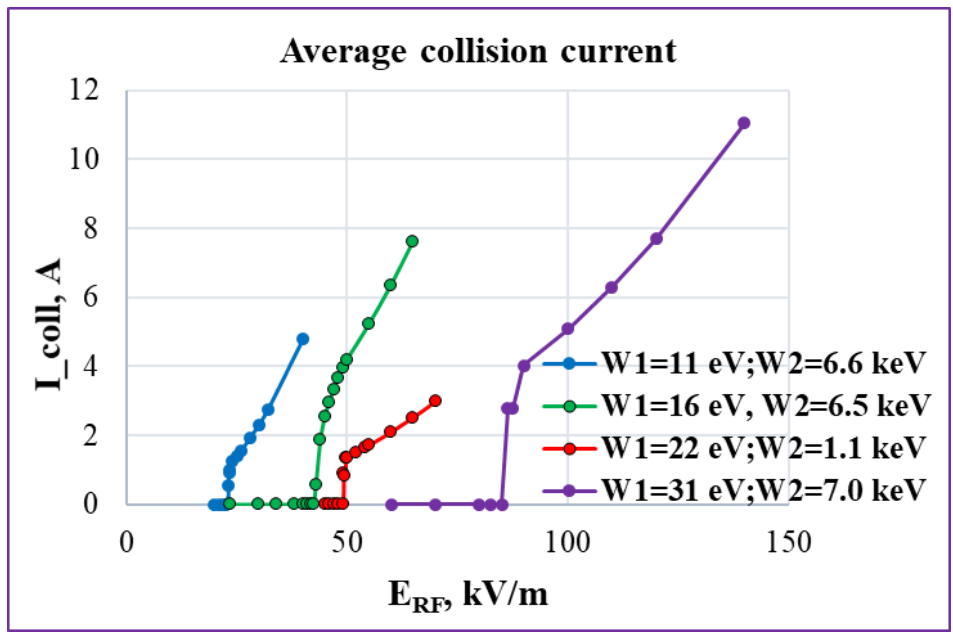

Figure 15: The collision currents vs RF field amplitude for different crossovers $W_1$ and $W_2$.

Fig.15 and 16 show the complete set of the collision currents and energies vs RF field amplitude. Apparently, the growth rates of the collision currents are correlated with second crossover $W_2$.

Both the threshold RF field amplitude and the collision energy at the thresholds are linear functions of first crossover $W_1$ as shown in Fig.17. The theoretical prediction of threshold made with formula (2) is also shown to compare with. There is a disagreement between the theory and the simulations, and it increases dramatically with increasing of $W_1$. The theory assumes a polyphase regime at low $E_{DC}$, and it assumes also that it remains polyphase. But the voltage (i.e. $E_{DC}$) sharply jumps to much higher level at threshold (Fig.18). It means that the electrons with higher initial energy $W_0$ and therefore more numerous become resonant, so the overall MP process becomes dominantly resonant.

The DC voltage shown in Fig 18 is relative as it has been mentioned earlier, because it is measured on the plate side without MP, but we can make some qualitative speculations about surface charge at saturation.

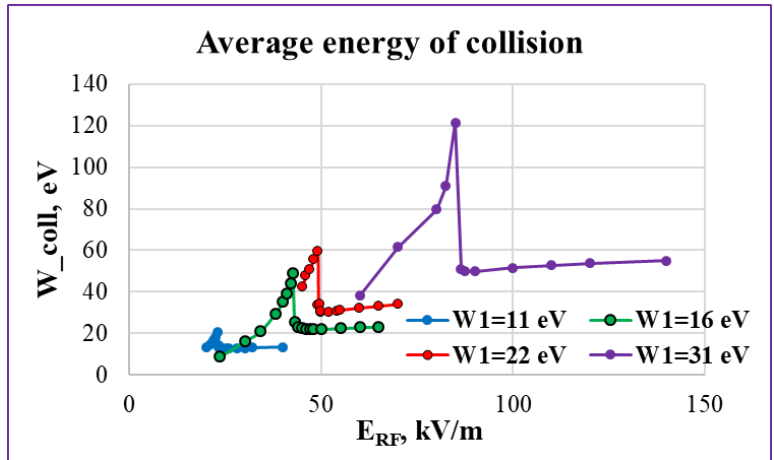

Figure 16: The collision energies vs RF field amplitude for different first crossovers $W_1$.

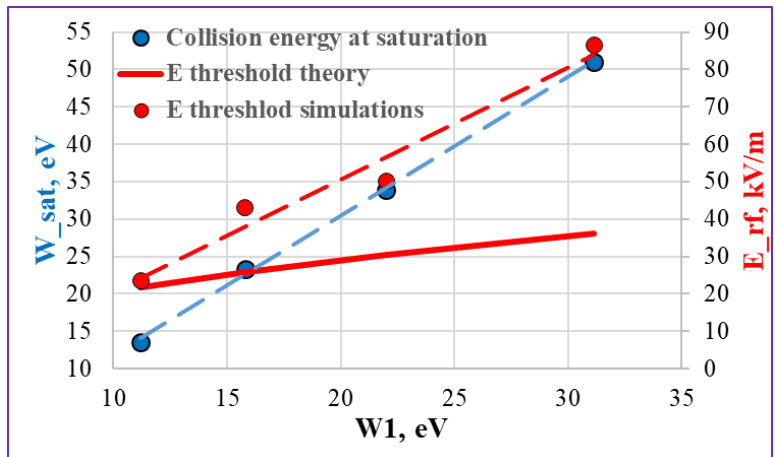

Figure 17: RF field threshold and average collision energy at saturation vs first crossover of SEY. The threshold according to theory [2], which assumes completely polyphase regime, is given for comparison.

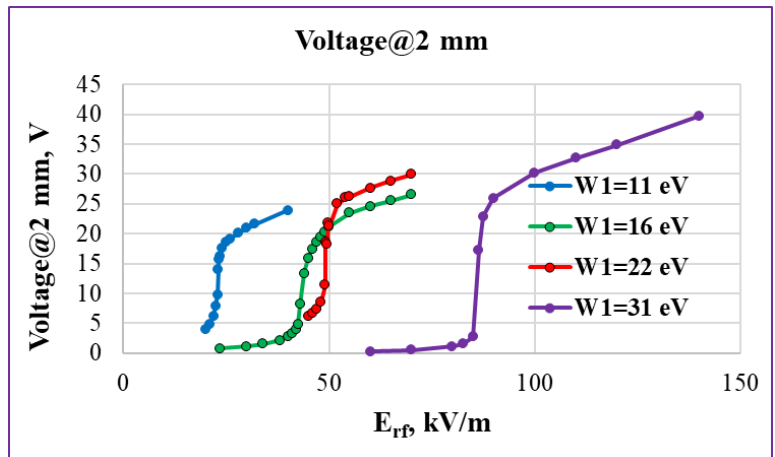

Figure 18: The voltage monitor readings vs RF field amplitude and different first crossovers of SEY functions.

Under assumption that the resonance MP dominates at saturation, the initial energy of resonant secondary electrons is one of the factors that regulates charging of dielectric. The number of resonant electrons among the secondaries should be high enough to support multipacting. In other words, the resonant electrons must have an initial energy hovering around $W_{max}$ of PDF. In simplified picture without considering other factors, if $E_{DC}$ increases and exceeds the level, above which the secondary electrons with initial energy higher than $W_{max}$ become resonant, then the number of such resonant particles goes down, a dielectric discharges and $E_{DC}$ returns to some equilibrium level.

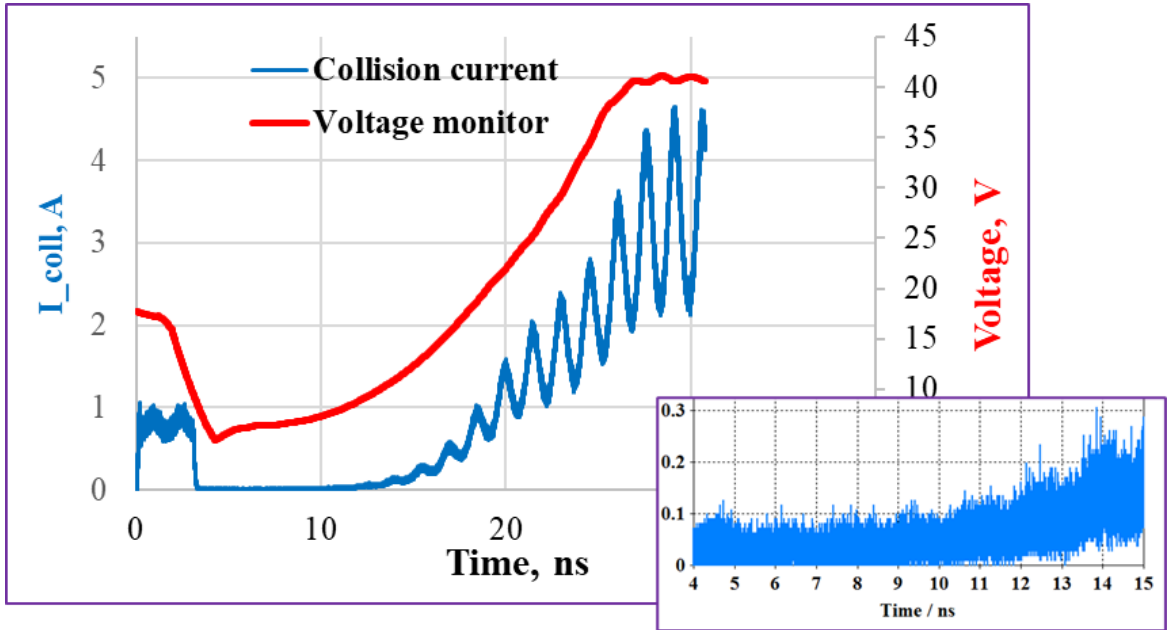

Figure 19: Transition of MP development from the polyphase regime to the resonance one.

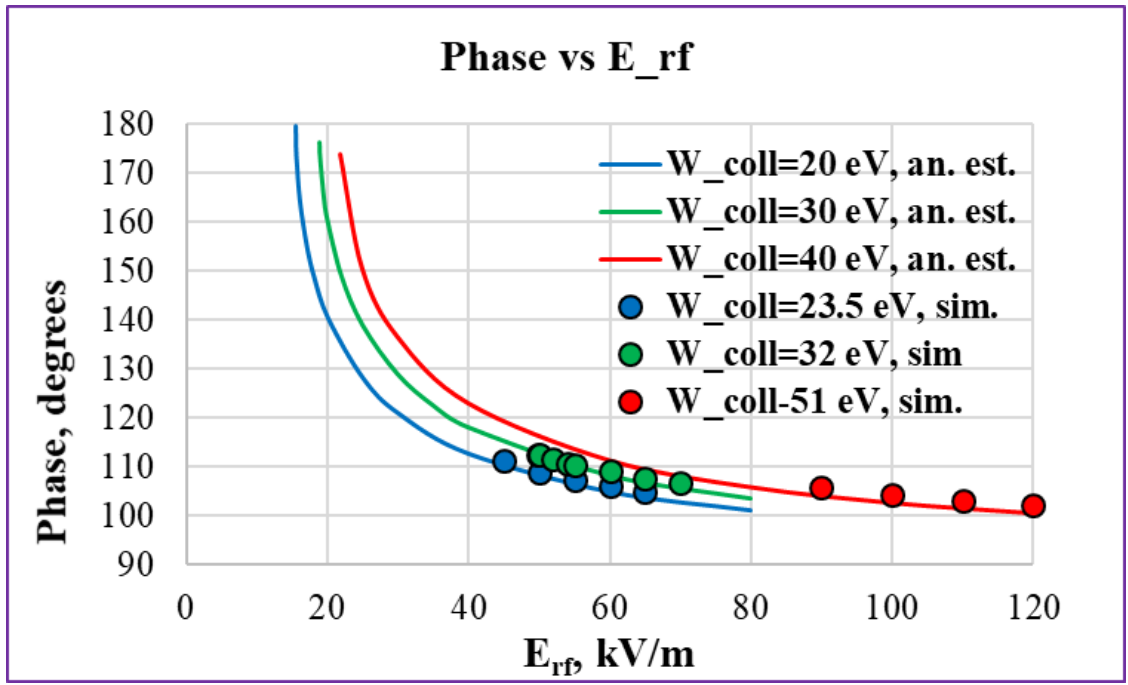

Figure 20: Average initial phase $\theta$ of the resonant electrons that provides certain W_coll vs E_rf. The phases calculated from the simulations are compared to the analytical calculations using (1).

Fig.19 shows the charging of dielectric plate that develops synchronously with increase and saturation of the collision current. The insert in the right corner of the figure shows almost uniform distribution of the collision current over time at low levels of electrostatic field in the beginning of MP development. This confirms the speculation that MP on a dielectric starts in a polyphase (non-resonant) regime, which gradually transforms to a dominantly resonance process as the charge on the dielectric increases.

The fact that the collision energy at saturation almost does not depend on RF field level can be explained by a variation of initial phase of resonant particle. Fig.20 shows the evaluation of the average initial phases that provide different fixed energy of collision based on the simulations and compared to the analytical calculations of the same energy for resonant particle (time of flight is T/2) and variable initial phase using (1).

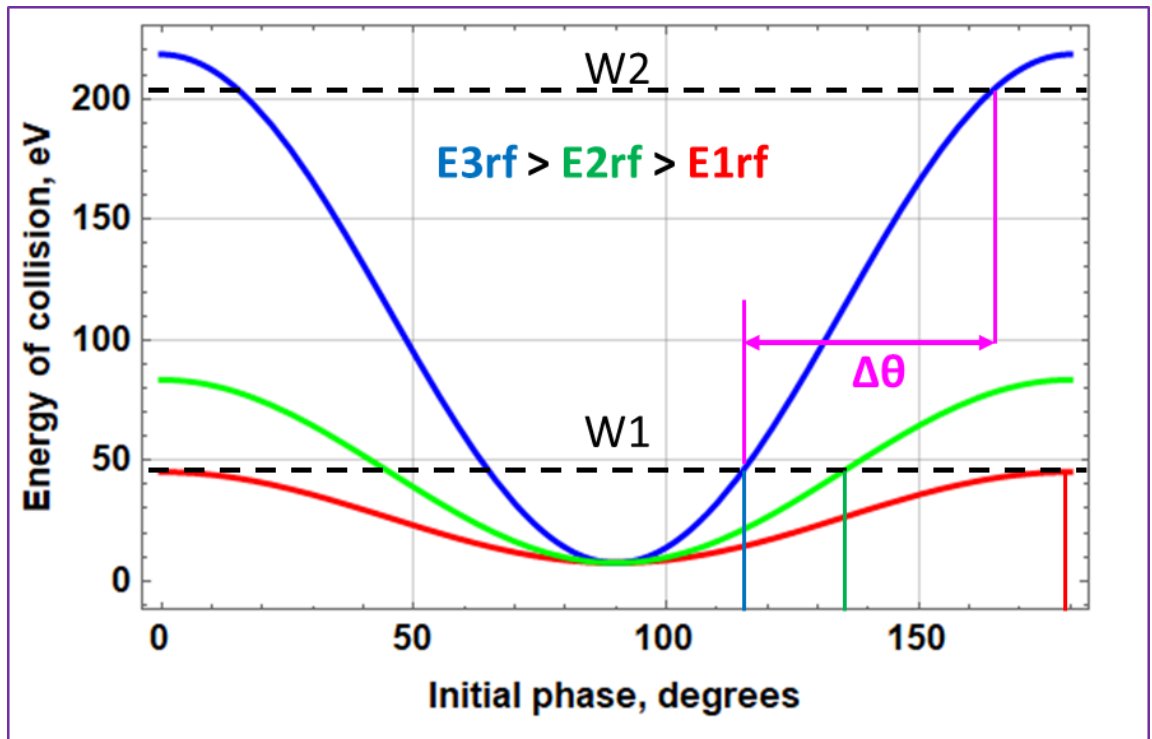

Figure 21: Analytical evaluation of the resonant particles' energy of collision as a function of initial phase at different amplitudes of RF field. The crossover energies $W_1$ and $W_2$ are arbitrary and serve for qualitative explanation of the initial phase variation.

The Fig.21 illustrates the mechanism of the initial phase variation of the resonant particles. The solid lines are energy of collision of the resonant particles vs their initial phases at given level of RF field. The collision energies of the particles were calculated using equations (1). The behavior of the energy of collision vs RF field level is explained by the fact that a resonant particle is accelerated during part of time of flight and deaccelerated during the rest part of flight, excluding initial phases 0° and 180° (or accelerated all the time, but in opposite directions). Therefore, the energy of collision is a difference between acquired and lost energies. The dashed lines are the crossover energies $W_1$ and $W_2$ of SEY, their levels are arbitrary and chosen to fit the plot conveniently. Formally MP starts at RF field level of E1rf and its initial phase of 180°, once the energy of collision reaches first crossover $W_1$ (the phases 90-180° were chosen for speculations because in this interval auto phasing and phase stability are expected). From this point of view, it is clear why the RF field level at which MP starts depends on $W_1$. But the number of resonant particles that emitted exactly at 180° is too small to support multiplication. The MP starts when the range of appropriate initial phases Δθ is big enough to develop multipacting process, say at field level E2rf. The multipactor continues with RF field increase as long as Δθ stays sufficiently big. With further RF field increasing the Δθ starts shrinking, and MP should stop when Δθ gets lower some critical value, though that level of RF field was not reached in the simulations.

## Summary

This work is to summarize the results of PIC simulations of one-side multipactor on dielectric and to accumulate the observations of the features of the process, some of which are banal, some are not understood in full yet. Therefore, the following list of the observations is rather a list of suppositions that require further study and verifications than something conclusive.

- At any DC field there is always a synchronous secondary electron with appropriate initial velocity due to continuous PDF.
- At saturation the charging process adjusts DC field in such a way that secondary electrons with initial energy equal or close enough to maximum of PDF are synchronous
- The surface charging occurs only if $W_{coll} > W_1$ at least for some part of secondary electrons, i.e. $E_{RF}$ must be sufficient to accelerate electrons above this energy.
- There are two stage of MP development: polyphase regime in the beginning of MP (mostly) and resonant regime at saturation (mostly). The MP starts at lower levels of RF field than polyphase theory predicts. So, apparently the resonant electrons also contribute to the MP development at the early stages.
- At saturation $E_{DC}$ and $W_{coll}$ vs $E_{RF}$ are approximately constant for given SEY: the crossovers are important defining parameters, especially $W_1$. Maybe PDF also plays its role, but it was not a variable in this work.
- MP dynamic at saturation has tendency to establish average energy of collision $W_{coll}$ close to the first crossover $W_1$ of SEY.
- MP dynamic via synchronous phase keeps average energy of collision $W_{coll}$ constant while field $E_{RF}$ is varying.
- Collision $I_{coll}$ at saturation increases with increase of $E_{RF}$, supposedly due to increase of the range of acceptable phases $\Delta\theta$. Unfortunately, it is not clear what happens at very high RF field because of numerical simulation difficulties. Gradient $dI_{coll}/dE_{RF}$ is correlated with second crossover $W_2$.
- The horizontal and vertical movements appear to be independent according to the uncoupled equations (1). But the same charge simultaneously affects induced electrostatic field $E_{DC}$ and produces phase-dispersing effect in horizontal direction [4]. Therefore, the space charge in some form should appear in both equations (1) and make them coupled.

# Appendix

## One side multipactor on dielectric at different frequencies

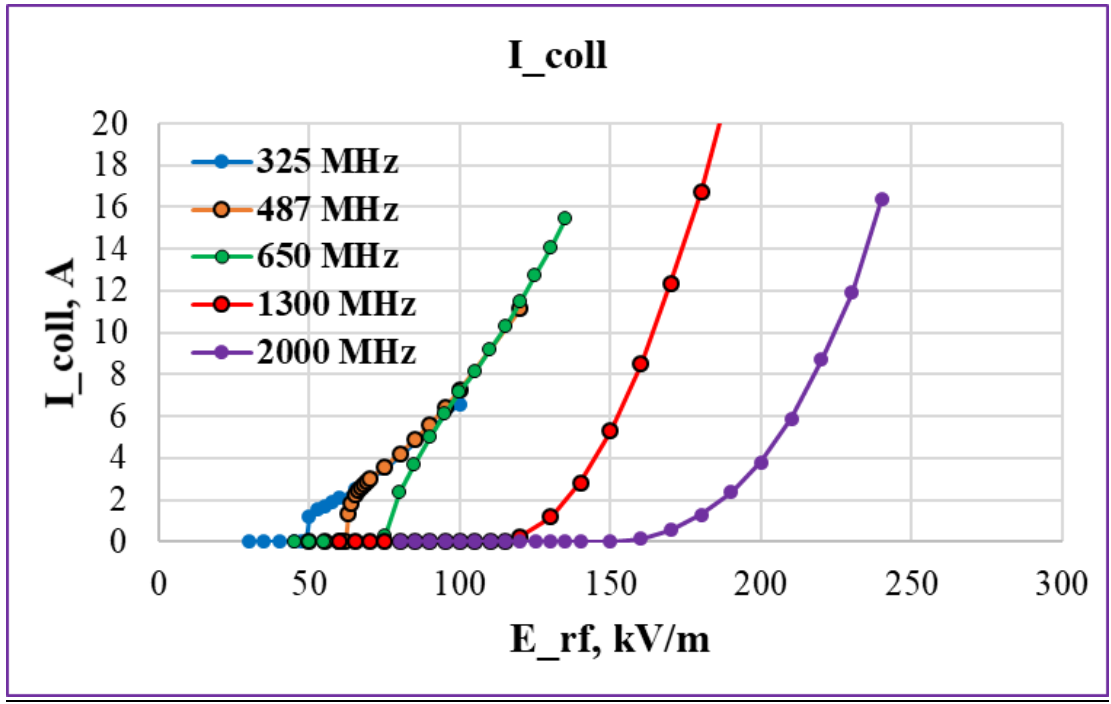

Figure 22: Shift of the RF electric field thresholds for different frequencies

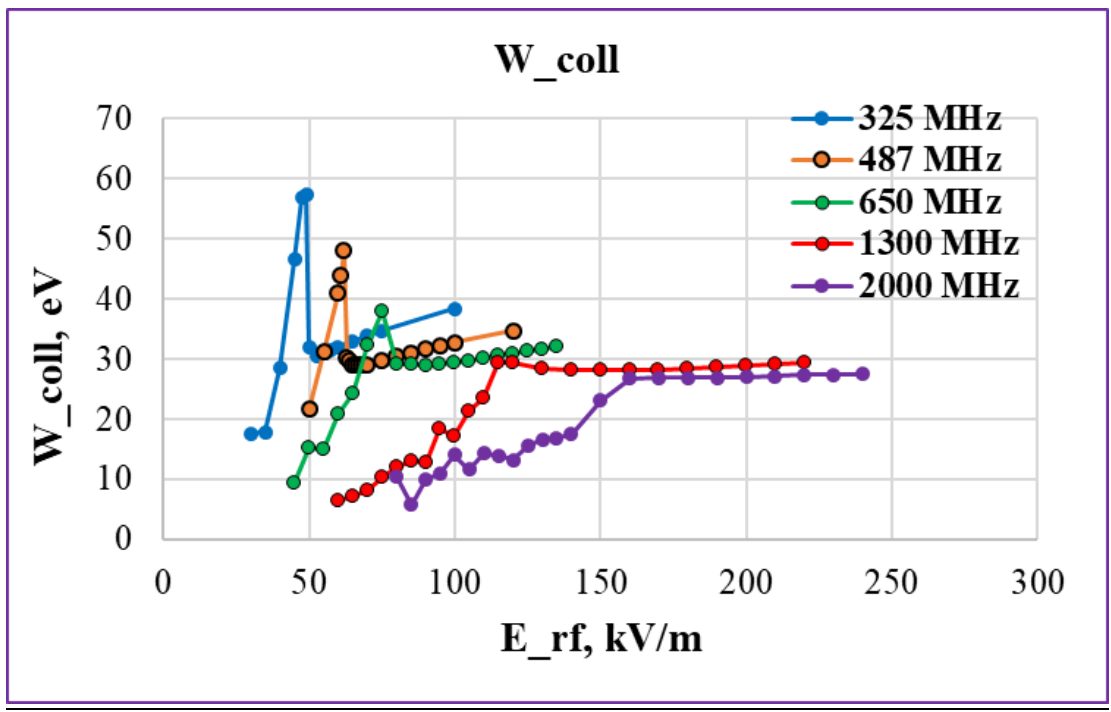

Figure 23: Collision energies vs RF field strength for different frequencies.

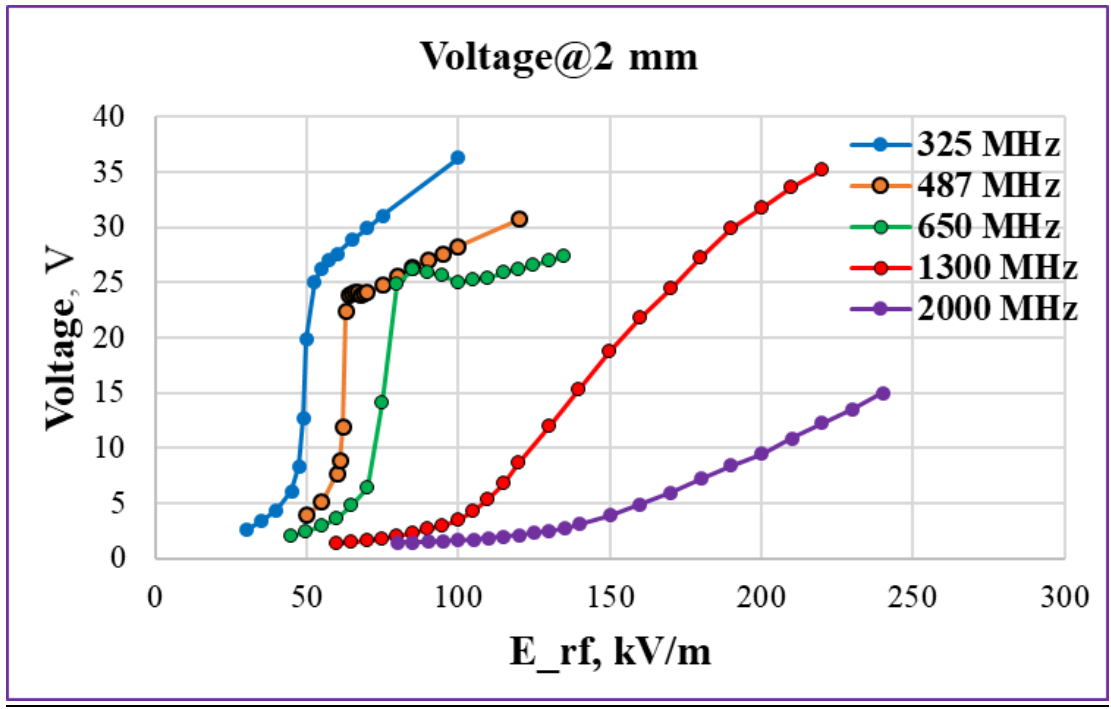

Figure 24: Voltage monitor reading vs RF field strength for different frequencies. Data for high fields may be inaccurate, since the simulations with higher frequencies are more demanding.